# Electromagnons, magnons, and phonons in Eu$_{1-x}$Ho$_x$MnO$_3$

Zhenyu Chen[1], M. Schmidt[1], Zhe Wang[1], F. Mayr[1], J. Deisenhofer[1], A. A. Mukhin[2], A. M. Balbashov[3], and A. Loidl[1]

[1] Experimental Physics V, Center for Electronic Correlations and Magnetism, University of Augsburg, 86135 Augsburg, Germany
[2] Prokhorov General Physics Institute, Russian Academy of Sciences, Moscow, 119991 Russia
[3] Moscow Power Engineering Institute, ul. Krasnokazarmennaya 14, Moscow, 111250 Russia

(18.01.2016)

**Abstract**

Here we present a detailed study of the THz and FIR response of the mixed perovskite manganite system Eu$_{1-x}$Ho$_x$MnO$_3$ for holmium concentrations x = 0.1 and 0.3. We compare the magnetic excitations of the four different magnetically ordered phases (A-type antiferromagnetic, sinusoidally modulated collinear, helical phases with spin planes perpendicular to the crystallographic **a** and **c** axes). The transition between the two latter phases goes hand in hand with a switching of the ferroelectric polarization from **P**||**a** to **P**||**c**. Special emphasis is paid to the temperature dependence of the excitations at this transition. We find a significant change of intensity indicating that the exchange striction mechanism cannot be the only mechanism to induce dipolar weight to spin-wave excitations. We also focus on excitations within the incommensurate collinear antiferromagnetic phase and find a so far unobserved excitation close to 40 cm$^{-1}$. A detailed analysis of optical weight gives a further unexpected result: In the multiferroic phase with **P**||**c** all the spectral weight of the electromagnons comes from the lowest phonon mode. However, for the phase with the polarization **P**||**a** additional spectral weight must be transferred from higher frequencies.

**I. Introduction**

The giant cross coupling of magnetization **M** and polarization **P** with electric and magnetic fields in spin-driven multiferroics opened new perspectives in the field of multiferroics, which therefore makes them interesting for basic research and promising for potential applications. Already in the sixties Smolenskii and Ioffe[1] searched for simultaneous ordering of electric dipoles and localized magnetic moments in $ABO_3$ perovskites. Their idea was a combination of lone-pair ions at the A site, favoring ferroelectricity, and ions with partly filled d-shells at the B site, responsible for magnetic order. The enormous variety of possible technological applications has early on been outlined by Wood and Austin.[2] Scientific work devoted to the coexistence of magnetic and polar order in bulk materials, combined with possible magnetoelectric effects gained considerable impetus specifically after the discovery of spin-driven ferroelectricity in perovskite manganites.[3] The appearance of polarization in these materials can well be explained by a spin-current model[4] or by the inverse Dzyaloshinskii-Moriya interaction.[5, 6] According to these mechanisms, the direction of the polarization can be calculated from $\vec{P} \propto \vec{e}_{ij} \times (\vec{S}_i \times \vec{S}_j)$, where $\vec{e}_{ij}$ is the vector connecting the adjacent spins $\vec{S}_i$ and $\vec{S}_j$. Later, two additional mechanisms have been found which drive the ferroelectricity by magnetic order: exchange-striction[7] and spin-dependent *p-d* hybridization.[8,9] Exchange striction has also been identified as the main microscopic mechanism to establish ferroelectricity in all magnetic phases[10] in the multiferroic skyrmion host $GaV_4S_8$.[11] There exist a number of reviews on multiferroics documenting the enormous amount of experimental evidence and theoretical knowledge that has been accumulated on this new class of materials during the last decade.[12,13,14,15,16,17]

Electromagnons (EMs) are the generic excitations of magnetoelectrics. The existence of hybrid ferroelectric and ferromagnetic excitations was theoretically proposed almost 5 decades ago[18, 19] and has been interpreted in a polariton-like picture.[20] These coupled excitations have originally been termed Seignette Magnons[19] and later have been renamed Ferroelectromagnons.[18] The first experimental observation of this new class of excitations has been reported in the multiferroic perovskites $TbMnO_3$ and $GdMnO_3$.[21] The authors of Ref. 21 termed these excitations Electromagnons, which nowadays appears to be the generally accepted name for generic magneto-electric excitations in polar magnets. The dynamics of spin-driven ferroelectrics characterized by spin waves coupled to the electric polarization has been calculated by Katsura *et al*.[22] in good agreement with experimental observation.

Soon after, further identification of electromagnon response has been made in multiferroic $YMn_2O_5$ and $TbMn_2O_5$.[23] Colossal magnon-phonon coupling and the characteristic twin-peak structure of EMs in multiferroic perovskites, with two excitations close to 25 and 75 cm$^{-1}$, has been documented for $Eu_{0.75}Y_{0.25}MnO_3$.[24] Here, for the first time it has been noticed that in multiferroic perovskites, while the static polarization is strictly perpendicular to the chirality and the propagation vector of the spiral as outlined by the spin current-mechanism[4,5,6] the electromagnon response obviously is not, but seems to be pinned to the lattice. In perovskite manganites EMs are excited only with the ac electric field **E**$^\omega$ parallel to the crystallographic **a** direction, **E**$^\omega$||**a**.[24,25] Later on it has been shown that the dominant electromagnon response in the multiferroic manganites stems from a strong and dominating symmetric Heisenberg exchange[26] and, hence, symmetric superexchange contributions are responsible for the dynamic magnetoelectric effects which from now on will be termed exchange-striction mechanism.

Meanwhile, electromagnons have been found in several other compounds[27, 28, 29, 30, 31, 32] and there are some reports about electromagnons found by Raman scattering.[33, 34] Recently, the dynamic response of magnetoelectric materials gained a lot of attention, as especially the directional dichroism at electromagnon frequencies provides a high application potential, for example in directional light switches.[35][36][37] In addition, it was demonstrated that atomic-scale magnetic structures can be directly manipulated by THz light with electromagnon frequencies.[38]

In this work we investigated the low energy excitations of $Eu_{1-x}Ho_xMnO_3$ by THz and FIR spectroscopy. This system has been characterized by dielectric and magnetic measurements[39,40] establishing a detailed phase diagram with a regime close to x = 0.3 where the polarization switches from **P||a** to **P||c** as function of temperature. In Ref. 39 it has been argued that this polarization switching as function of Ho concentration and temperature is driven by the strong magnetic anisotropy introduced by the Holmium ions. The competition between the two orientations of the basal plane of the spin cycloidals originates from the competition between single-ion anisotropy and the Dzyaloshinskii-Moriya interaction.[41] We measured the optical response of crystals with this concentration and compared it with results from crystals with holmium concentration x = 0.1, where the collinear sinusoidally modulated spin structure is followed by a canted antiferromagnetic (AFM) phase.

**II. Experimental Details**

Single crystals of $Eu_{1-x}Ho_xMnO_3$ have been grown in a concentration range 0 ≤ x ≤ 0.5 by a floating zone method with optical heating in argon atmosphere. Details are given in Ref. 39. Room temperature x-ray diffraction of powdered single crystals proved that all crystals grown have the correct *Pbnm* orthorhombic structure with no impurity phases above experimental uncertainty. All samples have been characterized by magnetic susceptibility, magnetization, heat capacity, and dielectric measurements.[39, 40] A phase diagram which has been slightly improved and completed in comparison to Ref. 39, is given in Fig. 1. Above approximately 50 K all samples are paramagnetic (PM) and reveal a sequence of phase transitions into an incommensurate (IC) collinear antiferromagnetic (AFM) phase at slightly lower temperatures. Finally, the ground state is the canonical A-type AFM structure of the orthorhombic perovskites for x < 0.15 or helicoidal antiferromagnetic phases for x > 0.15. Depending on the concentration x, the A-type AFM phases possibly reveal slight canting of the spin structure and we specify all these phases as A-type antiferromagnets. In the helicoidal phases spontaneous polarization, i.e. ferroelectricity, appears with the polarization parallel to the crystallographic **a** direction for x < 0.35 and parallel to the **c** direction for 0.35 < x ≤ 0.5. At low temperatures and for a narrow range of Holmium concentrations between 0.15 and 0.35, the polarization reveals a thermally induced switching from **P||c** to **P||a**. For Ho concentrations 0.3 < x < 0.35, polarization appears in the **a** as well as in the **c** direction, but only for temperatures above 8 K. Following the spin current model, the polarization **P||a** signals helical spin order with a spin rotation within the **ab** plane, while **P||c** follows from a **bc**-plane cycloid (for the exact orientation of the spin plane also the magnetic susceptibility data from Refs. 39 and 40 is needed). It is unclear if in this concentration or temperature region both helical phases coexist or if the cycloidal has projections along all three crystallographic axes. In Ref. 39 arguments have been put forth, that the latter assumption is correct. On the basis of this knowledge we indexed the different phases in Fig. 1. At the highest concentrations and below 4 K another AFM phase evolves, which can be identified from magnetization and magnetic susceptibility measurements. It corresponds to an ordering of the rare earth magnetic moments which appears in pure HoMnO$_3$ at 6 K.[42] However, so far nothing is known about the spin order and the possible polar ground state of this phase.

Time-domain THz transmission experiments were carried out using a TPS Spectra 3000 spectrometer from TeraView Ltd. for temperatures between 3.5 K and room temperature. Reflectivity experiments from 60 to 700 cm$^{-1}$ were performed utilizing a Bruker Fourier-transform IR spectrometer IFS 113v equipped with a He-flow cryostat (Cryovac). THz spectra were taken from 5 cm$^{-1}$ up to approximately 100 cm$^{-1}$. In some cases, at least at higher frequencies and for specific sample geometries, the transmission was rather low and we were only able to measure THz spectra up to 70 cm$^{-1}$. In these cases, to arrive at broadband spectra, we combined THz with FIR spectra. To do so we converted the THz spectra, where real and imaginary part can be measured independently via transmission and phase shift, into reflectivity data and combined these with the FIR results by normalizing the FIR results to the calculated (THz) reflectivity in a wave-number range from 60 cm$^{-1}$ < $\nu$ < 75 cm$^{-1}$. From this so obtained normalized reflectivity we recalculated the complex index of refraction via the Kramers-Kronig relation. The normalization by a factor of up to 1.2 was necessary as the polished surfaces of the single crystals were not ideal for reflectivity measurements.

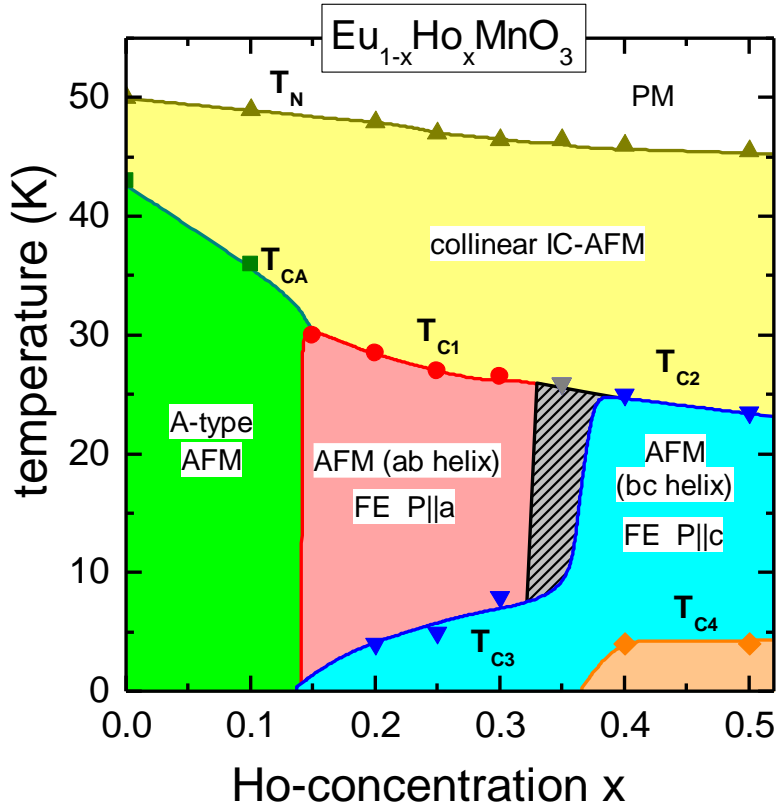

Fig. 1: (Color online) x,T Phase diagram of Eu$_{1-x}$Ho$_x$MnO$_3$. The paramagnetic (PM) high-temperature phase is followed by an incommensurate collinear antiferromagnetic phase (collinear IC-AFM), yellow) and at low temperatures, as function of Ho concentration by a sequence of A-type antiferromagnetic (A-type AFM, green) and helicoidal multiferroic phases (helical AFM, FE with **P**||**a**, (red) and **P**||**c** (blue). There is a further AFM phase below 4 K and beyond x = 0.4, probably with ordered rare earth moments (brown). The phase transition temperatures $T_N$, $T_{CA}$, and $T_{Ci}$ (i = 1 – 4) are indicated. In the shaded area coexisting ferroelectric polarization with **P**||**a** and **P**||**c** can be observed (see text).

## III. Experimental Results and Discussion

In this work we will focus on real and imaginary part of the complex index of refraction, $n^* = n + ik = \sqrt{\varepsilon\mu}$, as measured in $Eu_{1-x}Ho_xMnO_3$ as function of wave number and temperature for concentrations x = 0.1 and 0.3. For the analysis of the electromagnon or magnetic response, we will present mainly THz or - whenever necessary - combined THz and FIR results. The data have been analyzed assuming that the THz response is dominated by the complex dielectric constant and that the real part of the magnetic permeability $\mu'$ is close to 1 and the imaginary part, $\mu''$ negligibly small. This assumption is justified for the complete frequency range with the exception of the response close to the eigenfrequencies of antiferromagnetic resonances (AFM resonances) or electromagnons with a mixed magnetic and dipolar response. To acknowledge the magnetic contributions, in the following we always write $\varepsilon\mu$, although the calculation is the same as for the dielectric constant $\varepsilon$ of a non-magnetic compound.

As documented in Fig. 1, $Eu_{0.9}Ho_{0.1}MnO_3$ undergoes a phase transition from the paramagnetic into an IC collinear AFM phase at $T_N$ = 49 K. Based on the knowledge about the pure compound,[43] and mixed crystals from similar doping experiments,[44] as well as from the detailed neutron scattering work by Kenzelmann *et al.*[45] on $TbMnO_3$, we conclude that in this phase the manganese spins reveal a collinear spin structure which is sinusoidally modulated along the crystallographic **b** direction. The modulated spins are ferromagnetically aligned within the ab plane and are antiferromagnetically coupled along **c**. On further cooling the sample undergoes a magnetic phase transition into the A-type AFM phase at $T_{CA}$ = 36 K, the canonical antiferromagnetic structure of the orthorhombic perovskite manganites. In this phase the manganese spins reveal ferromagnetic order within the **ab** planes with the moments aligned along the crystallographic **b** direction. Neighboring **ab** planes are stacked antiferromagnetically along **c**. Any weak ferromagnetism due to canting would appear along the crystallographic **c** direction. For x = 0.3, the paramagnetic high-temperature phase is followed by the sinusoidally modulated collinear phase (IC- AFM) at $T_N$ = 47 K and subsequently at $T_{C1}$ = 26 K by a phase with helicoidal spin structure with the spin spiral within the **ab** plane. This phase is ferroelectric with the polarization **P||a**. Finally, at $T_{C3}$ = 8 K polarization switching is observed into a helicoidal phase with the ferroelectric polarization along the **c**-axis, **P||c**, characterized by a **bc** spiral of the manganese spins. So far the spin structures have not been determined by neutron diffraction but were deduced from magnetic susceptibility and polarization measurements[39,40] and by applying symmetry considerations from the spin current model. Similar systematic THz studies in mixed $Eu_{1-x}Y_xMnO_3$ compounds have been performed in Refs. 46 and 47.

**Magnon response for x = 0.1**

For x = 0.1 the paramagnetic phase is followed by a sinusoidally modulated collinear phase and below $T_{CA}$ by the A-type antiferromagnetic phase, characteristic for perovskite manganites (see Fig. 1). Both magnetic phases are not ferroelectric and hence not multiferroic, but according to Stenberg *et al.*[48] the low-frequency electromagnon should be visible in the sinusoidally modulated phase. In addition, magnetic dipole active antiferromagnetic resonance modes (AFM resonances) should easily be identified in THz transmission spectroscopy despite their very weak intensities. In the canted A-type antiferromagnetic phases of Sr doped $LaMnO_3$ detailed studies of AFM resonances have been performed by Ivannikov et al.[49] and by Mukhin et al.[50] Due to a canted spin structure at finite doping, these systems exhibit spontaneous ferromagnetic magnetization at low temperatures, and two modes, quasi-ferromagnetic (F) and quasi-antiferromagnetic (AF), corresponding to oscillations of the ferro- and antiferromagnetic vector were identified. AFM resonances in the canted A-type antiferromagnetic phase have also been studied in $Eu_{1-x}Y_xMnO_3$.[46,47] For the ac magnetic field applied along the **c**-axis, narrow and weak absorption modes were identified close to 20 cm$^{-1}$, corresponding to the quasi-AF mode of the A-type antiferromagnetic phase.

To detect possible electromagnon response in the sinusoidally modulated collinear phase, experiments on crystals with x = 0.1 seem to be very promising. For this concentration the collinear phase is followed by the canonical canted AFM structure where EMs certainly do not exist. At higher concentrations the paraelectric collinear phase is followed by the ferroelectric helical phase and precursor effects, i.e. electromagnon fluctuations can be expected, which makes the analysis more difficult. And indeed, in similar experiments in $Eu_{0.9}Y_{0.1}MnO_3$ with a similar sequence of phases as in the Holmium compounds investigated here, a significant increase in the real and imaginary part of the refractive index was detected in the collinear sinusoidally modulated phase which was interpreted as precursor effects of electromagnetic origin.[51]

In the course of this work polarization dependent measurements of the complex refractive index of $Eu_{0.9}Ho_{0.1}MnO_3$ have been performed on crystals cut perpendicular to the **a**-axis (**a**-cut), **b**-axis (**b**-cut), or **c**-axis (**c**-cut)

In Fig. 2 and 3 we plot results of THz transmission measurements in the **ac** plane (**b**-cut) of $Eu_{0.9}Ho_{0.1}MnO_3$. For these time domain experiments the sample thickness was 1.05 mm. The upper panel of Fig. 2 documents experiments with $H^\omega||\mathbf{a}$ and $E^\omega||\mathbf{c}$ showing the imaginary part of εμ at frequencies between 6 cm$^{-1}$ and 30 cm$^{-1}$ and temperatures between 4 K and 60 K. The lower panel shows the intensity contour plot of the imaginary part of εμ, as a color-coded plot of temperature (on a logarithmic scale) vs. wave number. We find a well-developed AFM resonance at 20.5 cm$^{-1}$ which exists in the low-temperature CA-AFM phase. The excitation is rather weak, but at the lowest temperatures exhibits a narrow linewidth well below 3 cm$^{-1}$ (full width at half maximum). This magnetic excitation becomes soft when approaching the phase boundary of the A-type AFM phase at $T_{CA}$ = 36 K. According to the theory of antiferromagnetic resonances[52,53] the resonance frequency follows the temperature dependence of the product of Weiss exchange field and anisotropy field and hence will be primarily dominated by the temperature dependence of the sublattice magnetization. In addition, the excitation also becomes very broad and when approaching $T_{CA}$ and AFM resonances do not exist as well-defined excitations in the sinusoidally modulated collinear IC AFM phase. At 45 K, well below $T_N$ = 49 K, a resonance absorption can hardly be detected in the upper frame of Fig. 2.

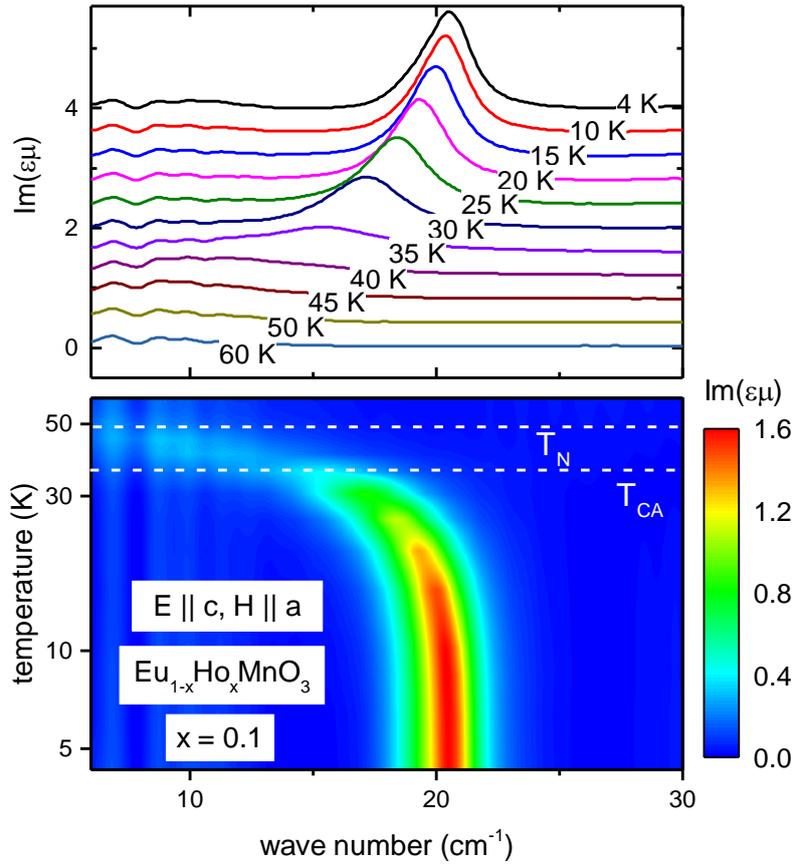

Fig. 2: (Color online) Imaginary part of εμ for Eu$_{0.9}$Ho$_{0.1}$MnO$_3$ within the crystallographic **ac** plane with polarization of the ac fields **H**$^ω$||**a** and **E**$^ω$||**c**. Upper frame: Selected scans for wave numbers between 6 cm$^{-1}$ and 30 cm$^{-1}$ and temperatures between 4 K and 60 K. The scan at 60 K is plotted at absolute values of the imaginary part of εμ. Subsequent temperatures are shifted each by Im (εμ) = 0.4 for clarity. At all temperatures the loss is close to 0 for frequencies outside the resonance conditions. Lower frame: Intensity contour plot where the imaginary part of the index of refraction as shown in the upper frame is color coded between 0 and 1.6 on a linear scale and plotted in a semi-logarithmic temperature wave-number plane. The magnetic ordering temperatures T$_N$ = 49 K and T$_{CA}$ = 36 K are indicated as dashed lines.

These experiments have been extended up to 70 cm$^{-1}$, and close to 55 cm$^{-1}$ a very weak excitation is observed with temperature independent eigenfrequency and intensity which goes to zero at approximately 60 K (not shown here). This excitation corresponds to a crystal-field excitation of the 4*f* states of the Ho ions and is very weak due to the low Ho concentration. We will discuss this excitation in more detail for x = 0.3.

With this **b**-cut crystal analogous experiments have been performed in the other polarization direction, with **H**$^ω$||**c** and **E**$^ω$||**a.** The results are shown in Fig. 3 and reveal a well-defined AFM resonance in the CA AFM phase at 20.5 cm$^{-1}$ behaving very similar to the observation for the other polarization direction **H**$^ω$||**a** and **E**$^ω$||**c** in Fig. 2.

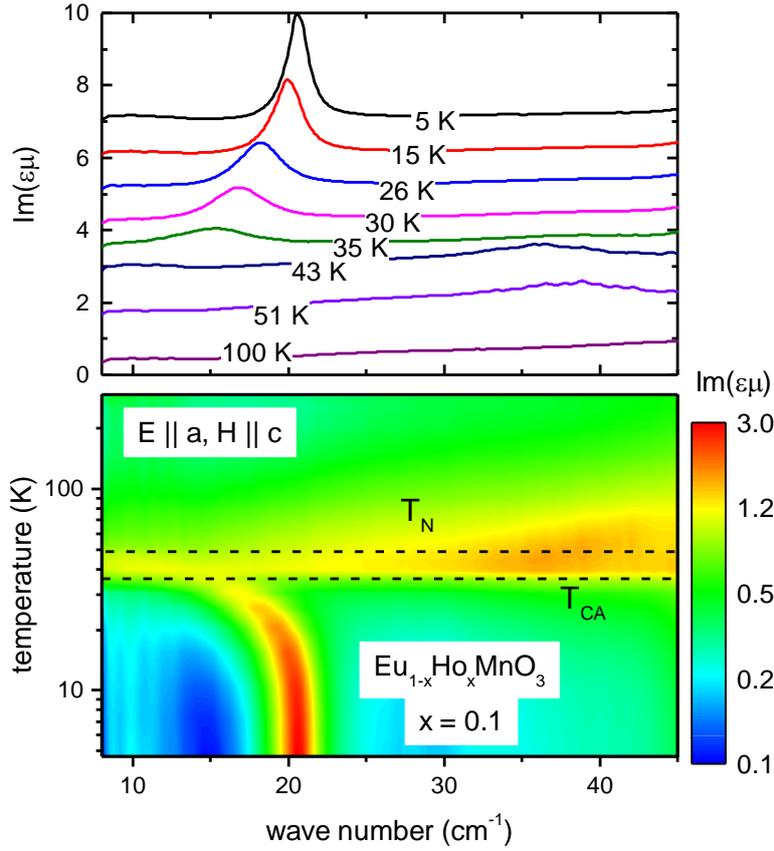

Fig. 3: (Color online) Imaginary part of εμ for $Eu_{0.9}Ho_{0.1}MnO_3$ within the crystallographic **ac** plane with polarizations of the ac fields $H^\omega||c$ and $E^\omega||a$. Upper frame: Selected scans at frequencies between 6 cm$^{-1}$ and 50 cm$^{-1}$ and temperatures between 4 K and 100 K. The scan at 100 K is plotted at absolute values of εμ. The scans at subsequent temperatures are shifted each by Im (εμ) = 1 for clarity. Lower frame: Intensity contour plot where the imaginary part of εμ is color coded between 0.1 and 3.0 in a logarithmic scale and plotted in a temperature vs. wave number plane in a semi-logarithmic representation. The magnetic ordering temperatures are indicated by dashed lines.

In addition, a weak and broad response shows up in the sinusoidally modulated AFM phase just below 40 cm$^{-1}$ which is best seen in the spectrum at 43 K in the upper frame of Fig. 3. The temperature and frequency regime of this broad feature is visualized in the color-coded plot in the lower frame. It abruptly appears at $T_{CA}$ and slightly extends into the paramagnetic phase. The nature and origin of this excitation are unclear at present. It could correspond to a damped AFM resonance or an EM-like response. Crystal field excitations can be excluded due to the fact that these modes only appear in the IC AFM phase. This mode is much weaker than an electromagnon excitation and much broader than an AFM resonance, but notably it only appears with the ac electric field $E^\omega||a$, which is the orientation sensitive to electric dipole excitations due to the exchange-striction mechanism in the manganites. Concerning a possible EM-like response of the collinear IC AFM phase, as theoretically predicted by Stenberg and de Sousa[48] only this 40 cm$^{-1}$ excitation seems to be a candidate. If this excitation corresponds to an electromagnon typical for the sinusoidally modulated IC AFM phase, it has to be explained why its eigenfrequency is enhanced by almost a factor of 2 compared to the low-frequency electromagnon of the multiferroic helicoidal phase (see below). It also could be an excitation driven by collinear exchange. In this case it is unclear why this excitation is unobservable in the other polarization direction in this plane. In any case, comparison with inelastic neutron scattering results in the IC AFM phase is highly desirable, which however to our knowledge has not been performed so far.

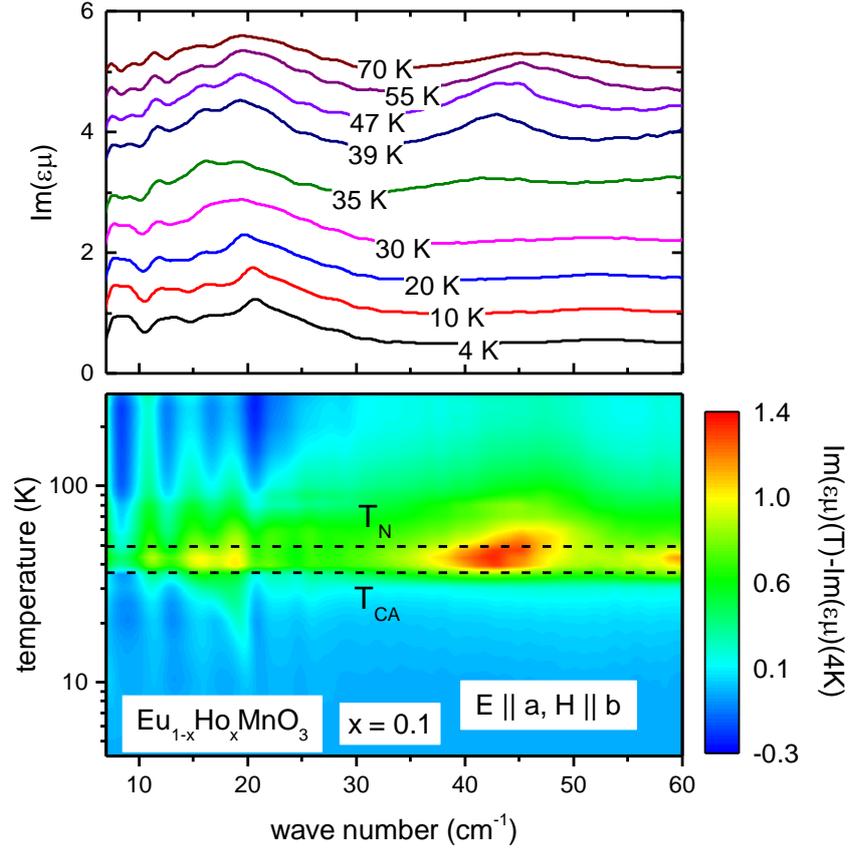

Fig. 4: (Color online) Imaginary part of εμ for Eu$_{0.9}$Ho$_{0.1}$MnO$_3$ within the crystallographic **ab** plane with polarizations of the ac fields **H**$^\omega$||**b** and **E**$^\omega$||**a**. Upper frame: Selected scans at frequencies between 6 cm$^{-1}$ and 60 cm$^{-1}$ and temperatures between 4 K and 70 K. The scan at 4 K is plotted at absolute values of εμ. The subsequent temperatures are shifted by Im (εμ) = 0.5 for clarity. Lower frame: Intensity contour plot where the imaginary part of the index of refraction subtracted by the values at lowest temperatures is color coded between -0.3 and 1.4 on a linear scale in a temperature vs. wave number plane. The magnetic ordering temperatures are indicated by dashed lines.

Fig. 4 shows the imaginary part of εμ as measured on a 760 μm thick sample within the crystallographic **ab** plane for the polarizations **H**$^\omega$||**b** and **E**$^\omega$||**a**. Similar to the previous measurement, one may also expect to detect electromagnons in this polarization. The upper panel shows representative scans for wave numbers from 6 cm$^{-1}$ to 60 cm$^{-1}$ and for temperatures between 4 K and 70 K. The lower panel documents an intensity contour plot shown in a temperature vs. wave number plane up to 300 K. To make temperature-dependent changes more visible, the 4 K data are subtracted from all the data.

We find no well-defined antiferromagnetic resonances at low temperatures in the CA-AFM phase, but only a very weak and broad response around 20 cm$^{-1}$ which however remains almost constant in intensity in both magnetic phases and exists even up to 70 K. Fig. 4 provides experimental evidence that the intensity of this excitation hardly changes between 4 K and 70 K and is completely independent of the magnetic state. Hence, it cannot be identified as an excitation of magnetic or magneto-electric origin. Similar to the observations documented in Fig. 3, a further excitation appears abruptly around 40 cm$^{-1}$ upon entering the non-collinear phase close to 36 K. On increasing temperatures this mode shifts towards higher frequencies and becomes smeared out in the paramagnetic phase. In addition, a steplike increase of Im ($\varepsilon\mu$) appears in this phase (Note that the frequency scans at each temperature are shifted for clarity by Im ($\varepsilon\mu$) = 0.5, but the increase of loss at 35 K is close to 1). It is frequency independent at least up to 60 cm$^{-1}$. We propose that this constant loss phenomenon indicates a continuum of low-frequency excitations in the longitudinal modulated IC AFM phase.

Experiments on the same sample, within the same frequency and temperature range have been performed for the polarization with **H**$^\omega$||**a** and **E**$^\omega$||**b** (not shown). We observed very similar results to the ones documented in Fig. 2 for ac fields **H**$^\omega$||**a** and **E**$^\omega$||**c**: a well-defined AFM resonance at 20.5 cm$^{-1}$ visible only in the CA-AFM phase, which is strongly damped when crossing into the sinusoidally modulated IC AFM phase, and in addition a weak crystal field excitation close to 55 cm$^{-1}$.

Summarizing the results for Eu$_{0.9}$Ho$_{0.1}$MnO$_3$ for all polarization directions, we found well defined AFM resonances at 20.5 cm$^{-1}$ in the canted antiferromagnetic structure for **H**$^\omega$||**a** and **H**$^\omega$||**c**, but no resonance for **H**$^\omega$||**b** irrespective of the orientation of the ac electric field. We identified a weak and rather broad excitation close to 40 cm$^{-1}$ in the IC collinear phase for **H**$^\omega$||**b** and **E**$^\omega$||**a** and for **H**$^\omega$||**c** and **E**$^\omega$||**a**.

**Electromagnons, antiferromagnetic resonances, crystal-field excitations and phonons for x = 0.3**

For Eu$_{0.7}$Ho$_{0.3}$MnO$_3$ we conducted combined THz and FIR experiments with wave numbers ranging from 5 cm$^{-1}$ up to 150 cm$^{-1}$ and for temperatures from 4 K up to room temperature to identify electric and magnetic dipole excitations. Measurements were performed within the **ab** plane with sample thicknesses of 315 µm for **E**$^\omega$||**a** and 770 µm for **E**$^\omega$||**b**. For measurements within the **ac** plane we used a sample thickness of 760 µm. As outlined above, in some cases we had to combine THz transmission with FIR reflectivity to arrive at broadband spectra. The low-frequency FIR spectra have been analyzed to calculate the temperature dependent oscillator strength of the lowest lying phonon and to determine the transfer of optical weight from this phonon to both electromagnons. The systematic investigation of samples with x = 0.3 seems highly interesting due to the fact that samples of this Ho concentration exhibit a temperature induced switching of the spin-spiral plane from the **ab** plane to the **bc** plane, with concomitant switching of the polarization from **P**||**a** to **P**||**c**. As documented in Fig. 1, the collinear AFM phase is established at T$_N$ = 47 K, followed by a transition into the multiferroic helicoidal phase with an **ab** helix and a polarization **P**||**a** at T$_{C1}$ = 26 K and finally at T$_{C3}$ = 8 K, the sample rotates the helicoidal basal plane into the **bc** plane with a polarization **P**||**c**.

As first step it seems interesting to turn the polarization of the light to the configuration **H**$^\omega$||**a**, **E**$^\omega$||**b**, where the responses of magnetic dipole-active spin excitations but no EM modes can be expected. Results for the imaginary part of $\varepsilon\mu$ can be seen in Fig. 5. In the upper panel we show the imaginary part of $\varepsilon\mu$ as function of frequency between 6 cm$^{-1}$ and 110 cm$^{-1}$ and for temperatures between 4 K and 70 K. In the lower panel there is a color-coded intensity contour map. For this polarization condition the transmission was high and THz experiments were analyzed up to 120 cm$^{-1}$. At low temperatures, in both helicoidal phases we find an excitation which we identify as AFM resonance at

20.5 cm$^{-1}$. It becomes soft, broad and smeared out when entering into the sinusoidally IC AFM phase. Two things are remarkable: i) The eigenfrequency of the AFM resonance in both low-temperature magnetic phases with helicoidal spin structure is at 20.5 cm$^{-1}$, the same as the eigenfrequency of the AFM resonance in the A-type AFM for x = 0.1. ii) At T$_{C3}$ = 8 K the system exhibits thermally induced polarization switching from **P**||**a** to **P**||**c**, indicating that the basal plane of the spin spiral turns from the **ab** into the **bc** plane. One would expect that AFM resonances with the polarization condition **H**$^\omega$||**a** are excited in the **bc** spiral and to a lesser degree in the **ab** spiral. Experimentally we observe a weaker but well defined and narrow resonance below T$_{C3}$, in the magnetic phase characterized by a **bc** spiral. The AFM resonance gets much broader in the magnetic phase with the **ab** spiral and gets heavily damped and soft in the collinear phase on further increasing temperatures. Also from the color coded plot one can see that the AFM resonance becomes weaker, but with a significantly smaller linewidth after switching the spiral plane from the **ab** into the **bc** direction.

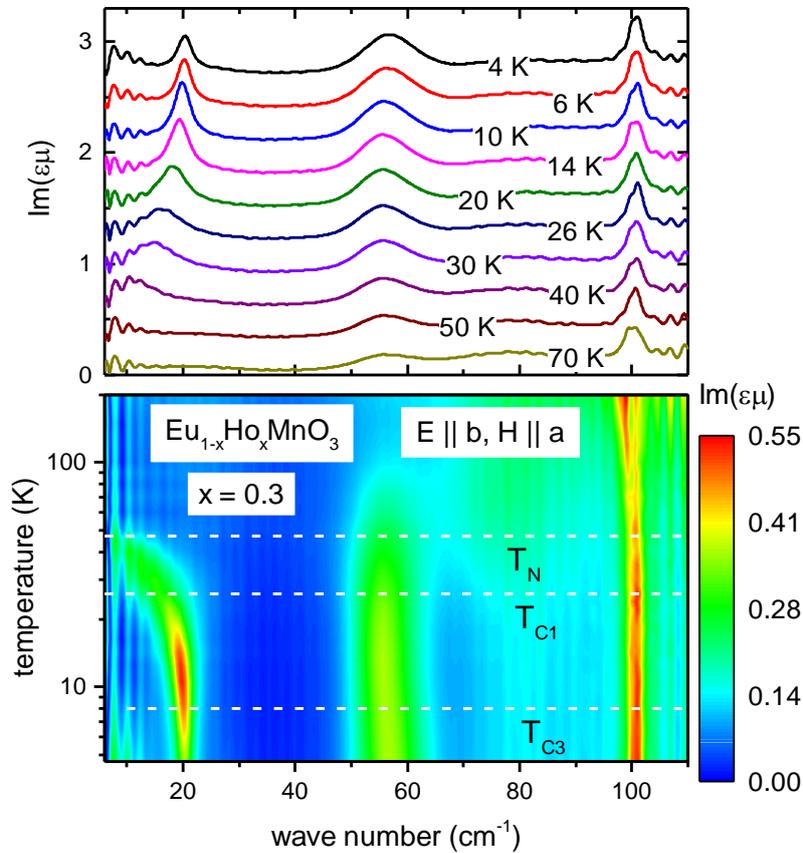

Fig. 5: (Color online) Imaginary part of εμ for Eu$_{0.7}$Ho$_{0.3}$MnO$_3$ within the crystallographic **ab** plane with polarizations of the ac fields **H**$^\omega$||**a** and **E**$^\omega$||**b**. Upper frame: Selected scans at frequencies between 6 cm$^{-1}$ and 110 cm$^{-1}$ and temperatures between 4 K and 70 K. The scan at 70 K is plotted at absolute values of εμ. The subsequent temperatures are each shifted by 0.3 for clarity. Lower frame: Intensity contour plot where the imaginary part of εμ is color coded between 0 and 0.55 and plotted in a temperature vs. wave number plane with a linear color code. The magnetic ordering temperatures, T$_N$ = 47 K, T$_{C1}$ = 26 K and T$_{C3}$ = 8 K, are indicated by dashed lines.

Further excitations can be observed at 55 cm$^{-1}$ and 100 cm$^{-1}$. The first is weakly temperature dependent and vanishes close to 70 K, and the second is almost temperature independent and exists up to room temperature. Both excitations exhibit no significant effects, neither in linewidth nor in eigenfrequency when passing the magnetic phase transitions. Following the observations in HoMn$_2$O$_5$,[54] we identify the excitation at 55 cm$^{-1}$ as a crystal field excitation of the 4$f$ electrons of Ho$^{3+}$. Ho$^{3+}$ with an $f^{10}$ electron configuration, according to Hund's rules, exhibits total spin **S** = 2 and total angular momentum **J** = 8. Even in cubic symmetry we expect 7 crystal field levels[55] and the excitation spectrum should be rather complex in orthorhombic symmetry. The intensity of crystal field excitations follows the difference in occupation numbers between the excited level and the ground state and with increasing temperature both levels should become more and more equally occupied yielding vanishing intensity of the light induced transitions. The temperature independence of the 100 cm$^{-1}$ mode, however, can hardly be explained along these lines, as it seems to survive up to room temperature which points rather towards a bosonic excitation. This 100 cm$^{-1}$ peak may be interpreted as a defect mode of phononic character.

Moreover, there is a weak absorption feature close to 80 cm$^{-1}$. This can be seen as well in the color-coded plot in the lower frame of Fig. 5. This is probably a crystal field transition from an excited level, because its intensity increases slightly with increasing temperature.

It is worth mentioning that similar THz experiments within the **bc** plane, with **E**$^\omega$||**b** and **H**$^\omega$||**c** and vice versa, gave no clear evidence for AFM resonance modes, neither in the collinear IC phase nor in both low-temperature helicoidal phases. There is no absorption close to 20 cm$^{-1}$ for **H**$^\omega$||**b** and only very weak absorption for **H**$^\omega$||**c**. In addition, the crystal field excitation appears for **E**$^\omega$||**b** at 60 cm$^{-1}$, but the sample was too thick to follow its temperature dependence.

Next we turn to the experiments in the **ab** plane with **E**$^\omega$||**a** where we expect the canonical EM response of the multiferroic manganites. Due to the significantly lower transmission and the given sample thickness we were not able to cover the complete frequency range up to 100 cm$^{-1}$ by THz spectroscopy alone, so we had to combine THz and FIR results as outlined above. In Figs. 6, 7 and 8 we combined THz and FIR results using THz data up to 60 cm$^{-1}$ and FIR reflectivity data at higher frequencies. The upper frame of Fig. 6 presents scans of the imaginary part of εμ for wave numbers between 8 cm$^{-1}$ and 95 cm$^{-1}$ and temperatures between 4 K and 70 K with polarized light with **H**$^\omega$||**b** and **E**$^\omega$||**a**. In multiferroic RMnO$_3$ electrically active magnon excitations have been observed always for **E**$^\omega$||**a**.[24, 25] Indeed, two electromagnons at 24 cm$^{-1}$ (LF-EM) and at 75 cm$^{-1}$ (HF-EM) appear just at the magnetic phase transition, $T_{C1}$ = 26 K, from the collinear sinusoidal phase into the low-temperature helicoidal spin state with ferroelectric polarization **P**||**a**. It is interesting to note that the LF-EM has a significantly smaller linewidth than the HF-EM, as found in all multiferroic perovskite manganites. Despite the fact that on further cooling the polarization switches into the **c** direction (**P**||**c**) at $T_{C3}$ = 8 K, the electromagnon response does not change significantly, which is the canonical observation in multiferroic perovskite manganites where electromagnons are tied to the lattice and are always excited with **E**$^\omega$||**a**.[24, 25] However, comparing the data at 4 K and 8 K, it is clear that the EM response specifically of the LF-EM becomes slightly weaker on decreasing temperatures indicating that the direction of the polarization and concomitantly the orientation of the spiral plane influence the EM intensity.

The detailed temperature dependence of this twin-peak response is documented in the color coded plot in the lower frame of Fig. 6. It is ideally suited to test experimental predictions that the LF-EM also exists in the IC AFM collinear phase[48] and to verify experimental results showing that the LF-EM also carries some weight due to the spin current mechanism, while the HF-EM is due to exchange striction and can only be observed in the helicoidal spin phase.[56] Some important observations can be immediately made:

I) Both excitations appear in the helical phase only. The slight increase in intensity for both excitations just above $T_{C1}$ probably stems from fluctuations when approaching the polar phase. From this it seems clear that both EMs cannot be observed in the collinear magnetic phase, or only with much weaker dipolar strength. One would expect intensity between $T_N$ and $T_{C1}$, which is not the case for either of the modes. However, the unusual response we observed in the IC collinear phase for x = 0.1 is approximately by a factor of 10 lower and could hardly be detected in these polarization conditions.

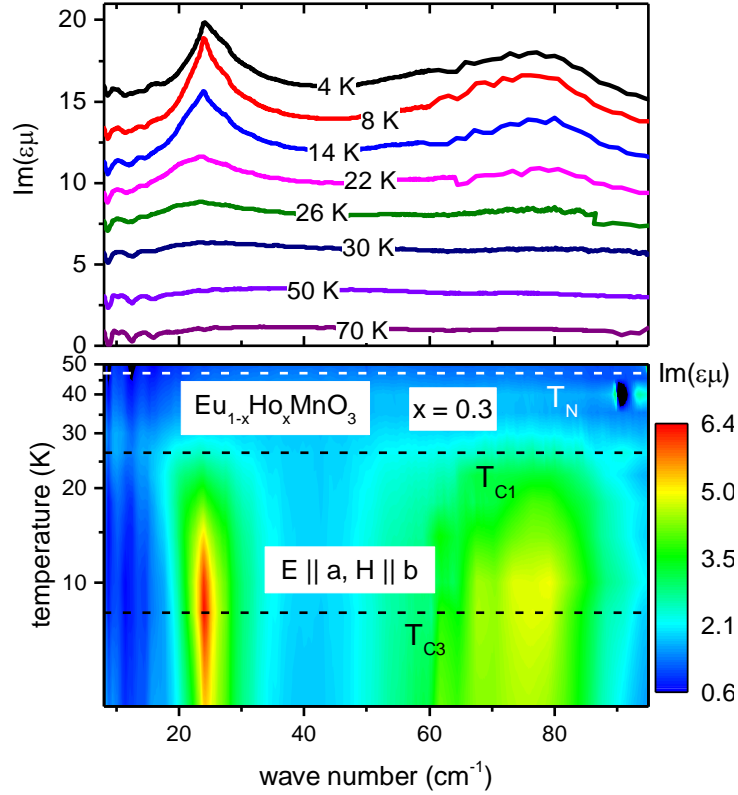

Fig. 6 : (Color online) Imaginary part of εµ for $Eu_{0.7}Ho_{0.3}MnO_3$ for wave numbers between 8 cm$^{-1}$ and 95 cm$^{-1}$ with $\mathbf{E}^\omega$||**a** and $\mathbf{H}^\omega$||**b** at a series of temperatures between 4 K and 70 K. For these measurements THz transmission and FIR reflectivity have been combined as outlined in the text. Upper frame: Scans at different temperatures which have been shifted by 2 for subsequent scans starting with T = 70 K. Lower frame: Color-coded plot of the loss on a logarithmic scale between 0.6 and 6.4. The magnetic ordering temperatures are indicated by dashed lines.

II) The lower frame of Fig. 6 also documents that the intensity of both EMs as a function of temperature is not as monotonous as one might expect. Both EMs seem to reach maximum intensity at the phase transition $T_{C3}$ to the low-temperature phase characterized by a polarization flop. A maximum in electromagnon intensity at 8 K is documented in the upper frame of Fig. 6 and can be also seen in the color coded plot in the lower frame.

To check the temperature dependence of the electromagnons systematically, we fitted both excitations by Lorentzian line shapes.

$$\varepsilon(\omega) = \varepsilon_\infty + \frac{\Delta\varepsilon\, \nu_0^2}{\nu_0^2 - \omega^2 - i\gamma\omega}$$

In Fig. 7 we show a prototypical example of fits of the real and imaginary part of εµ for $Eu_{0.7}Ho_{0.3}MnO_3$. The general agreement between fits and experimental results is not excellent but satisfactory and certainly allows to estimate the temperature dependence of the parameters of the Lorentzian fits, i. e. eigenfrequency $\nu_o$, damping $\gamma$, and dielectric strength $\Delta\varepsilon$. For the example the Lorentzian fits shown in Fig. 7 at 8 K result in the following parameters: $\nu_o = 25$ cm$^{-1}$, $\gamma = 9$ cm$^{-1}$ and $\Delta\varepsilon = 1.65$ for the LF-EM, while the HF-EM is best described by $\nu_o = 76$ cm$^{-1}$, $\gamma = 33$ cm$^{-1}$ and $\Delta\varepsilon = 1.4$. These results are characteristic of the twin-peak EM response in multiferroic manganites, with characteristic eigenfrequencies typical for all manganites, a similar dipolar strength, and a damping which differs almost by a factor of 4. Both excitations show maximal intensity close to 8 K, the temperature where polarization switching occurs. The dielectric strength decreases towards high and low temperatures. When approaching the phase transition to the collinear IC AFM phase both excitations broaden, significantly loose dipolar weight and become almost unobservable.

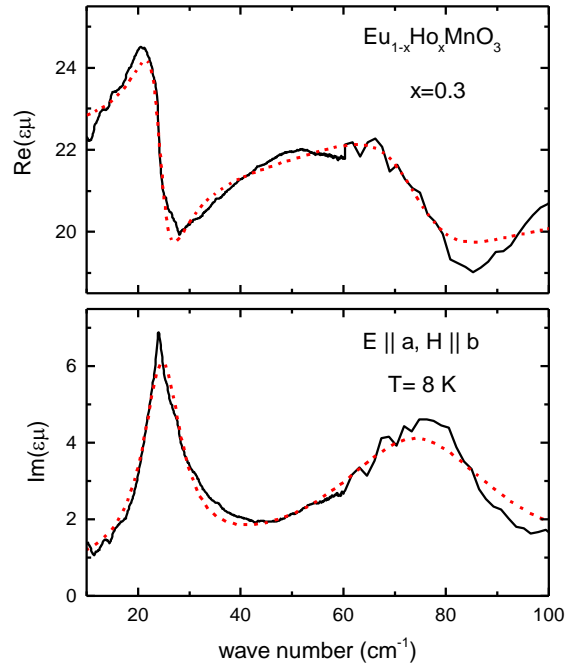

Fig. 7: (Color online) Real (upper frame) and imaginary part (lower frame) of $\varepsilon\mu$ for $Eu_{0.7}Ho_{0.3}MnO_3$ for wave numbers between 10 cm$^{-1}$ and 100 cm$^{-1}$ at 8 K (black solid lines) fitted by two Lorentzians (red dotted lines).

At this point it seems noteworthy to compare the parameters of EMs with those observed from AFM resonances as documented in Fig. 2 and Fig. 4 for Holmium concentrations x = 0.1 and in Fig. 5 for x = 0.3. The excitations documented in Figs. 2 and 4 result from spin waves in the canonical A-type AFM, the AFM resonance in Fig. 5 from spin waves in the helicoidal phases. It seems clear that the EM response documented in Fig. 6 is related to the same spin waves as the AFM resonance shown in Fig. 5. Nevertheless, in addition to the fact that the intensity of the electromagnon response is at least by one order of magnitude higher due to the electric dipole character of these excitations, the eigenfrequency is higher and the linewidth significantly broader. Even for the LF-electromagnon the linewidth is at least by a factor of 2 – 3 higher than for the corresponding AFM resonance. The width of the HF-EM is extremely broad and of order of half the eigenfrequency. This is the standard observation in all multiferroic manganites investigated so far and still needs to be explained.

Existing experimental evidence indicates that both electromagnons in perovskite manganites are pinned to the lattice and appear for $\mathbf{E}^\omega || \mathbf{a}$ only. However, it seems that non-negligible intensities probably result from spin current terms which should only be active in the high-temperature helicoidal phase with **P**||**c**. The maximum EM response appears at 8 K, just at the thermally induced switching from **P**||**a** to **P**||**c**. The moderate decrease of dipolar strength of both excitations towards low temperatures in the helicoidal phase with a **bc** spiral indicates that the electromagnon intensity is not totally independent of the orientation of the spiral plane. Exchange striction is the dominating contribution to the dynamic response, however, the spin current mechanism which is responsible for the static polar order, is obviously not fully negligible even for the dynamics. The decrease of intensity of the LF-EM of approximately 10 % on entering the low-temperature magnetic phase indicates that the exchange striction mechanism is at least by an order of magnitude stronger than the spin current mechanism. For the LF-EM in TbMnO$_3$, it was found that the spin current response is by a factor of 40 weaker than the exchange striction contributions.[56] From this observation we conclude that, while Heisenberg exchange is the dominant contribution to EM response, there seems to be additional weight due to spin-current terms. Figs. 6 and 7 also indicate that for this sample, there is no experimental evidence for a low-frequency satellite of the 20 cm$^{-1}$ EM as observed in TbMnO$_3$ and GdMnO$_3$.[57, 58]

Finally, we discuss the spin-lattice coupling in the multiferroic manganites. Numerous experimental evidence exists, documenting the coupling of electromagnons to phonons.[59, 60, 61] To follow this coupling in more detail, Fig. 8 shows the wave number range from 50 to 150 cm$^{-1}$, including the regime of the high-frequency electromagnon and the lowest frequency phonon mode. Some characteristic scans of the imaginary part of εμ vs. wave number are shown in the upper frame of Fig. 8. It is clear that the phonon mode dramatically loses intensity with the temperature dependent appearance of the electromagnon and that optical weight is shifted from the phonon to the EMs. This is elucidated in full detail in the lower frame of Fig. 8 using a color-coded intensity plot in a temperature wave number plane.

To get a more quantitative analysis of the coupling of electromagnons (or spin waves) to phonons, we calculated the spectral weight S in the low-frequency regime up to wave numbers of the first IR active phonon. The results are shown in Fig. 9. The spectral weight in optical spectroscopy is defined as the area under the real part of the conductivity and is given by S = $\varepsilon_o \int Im[\varepsilon(\omega)]\omega \, d\omega$, where $\varepsilon_o$ is the permittivity of free space. Focusing on the integral in the paramagnetic phase (50 K and 70 K), we find below 110 cm$^{-1}$ an increase of optical weight from 50 K to 70 K, indicating strong magnetic fluctuations already in the paramagnetic phase. The step like increase in spectral weight S by approximately 3 x 10$^5$ $\Omega^{-1}$cm$^{-2}$ at 110 cm$^{-1}$ corresponds to the oscillator strength of the low-energy phonon mode. On decreasing temperatures the optical weight of the electromagnons evolves now on top of this continuous background. Two smeared out steps in the frequency dependent optical weight close to 25 and 80 cm$^{-1}$ correspond to the LF-EM and HF-EM. The optical weight of the LF-EM is of order 1 x 10$^5$ $\Omega^{-1}$cm$^{-2}$ while S is approximately 4.5 x 10$^5$ $\Omega^{-1}$cm$^{-2}$ for the HF-EM. At the same time the optical weight of the phonon is significantly reduced, indicating that a large fraction of dipolar strength is transferred from the lowest optical phonon to the electromagnons. However, two important facts become obvious: The electromagnons carry the highest weight close to 8 K, where thermally induced switching of the polarization takes place and S becomes significantly reduced at 4 K with the largest contribution from the HF-EM. Interestingly, the 4 K curve almost reaches the 50 K curve directly after the frequency

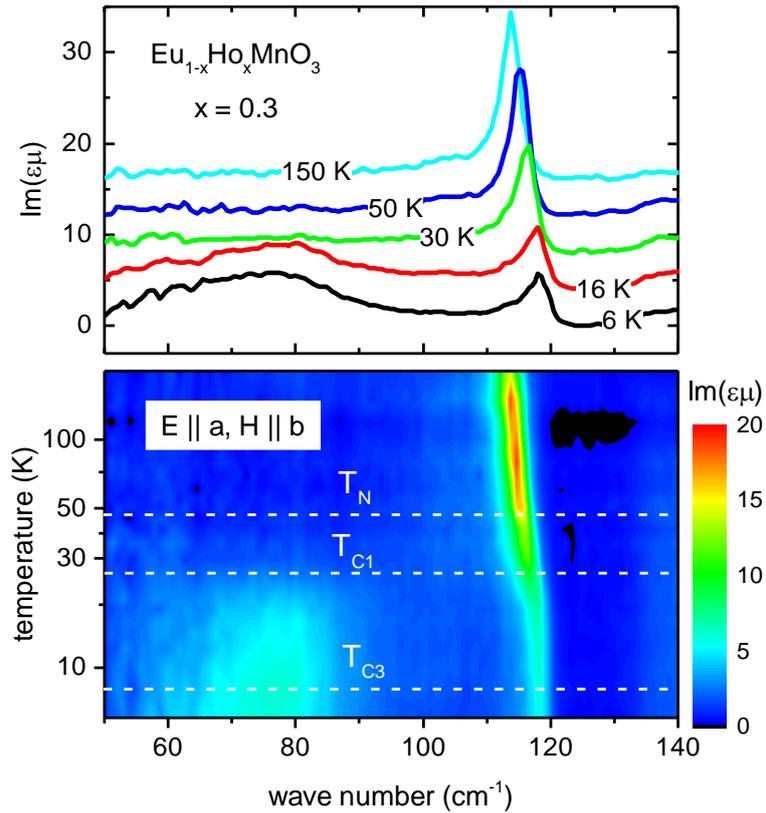

Fig. 8: (Color online) Upper frame: Imaginary part of εμ vs. wave number for **E**$^{\omega}$||**a** and **H**$^{\omega}$||**b** in Eu$_{0.7}$Ho$_{0.3}$MnO$_3$ at a series of temperatures between 6 K and 150 K. Lower frame: Color-coded plot of the intensity of the imaginary part of εμ in a temperature wave number plane for Eu$_{0.7}$Ho$_{0.3}$MnO$_3$. The color code is on a linear scale between 0 and 20. The characteristic magnetic transition temperatures are indicated as dashed lines.

of the lowest phonon mode. In this case, in the helicoidal AFM phase with **P**||**c**, all spectral weight of

the electromagnon is transferred from the lowest phonon. However, for 8 K the optical weight at 140 cm$^{-1}$ is far above the optical weight at 50 K, indicating that here significant optical weight stems from other frequency ranges.

In the inset of Fig. 9 we compare the optical weight determined from the electromagnons' response (8 – 105 cm$^{-1}$) with S as determined from the lowest phonon mode (105 – 120 cm$^{-1}$). We see that the EM response drops below 8 K, but remains finite even in the paramagnetic phase, indicating magnetic fluctuations. The phonon response increases at both magnetic phase boundaries $T_{C1}$ and $T_{C3}$ and continuously keeps growing even in the paramagnetic phase.

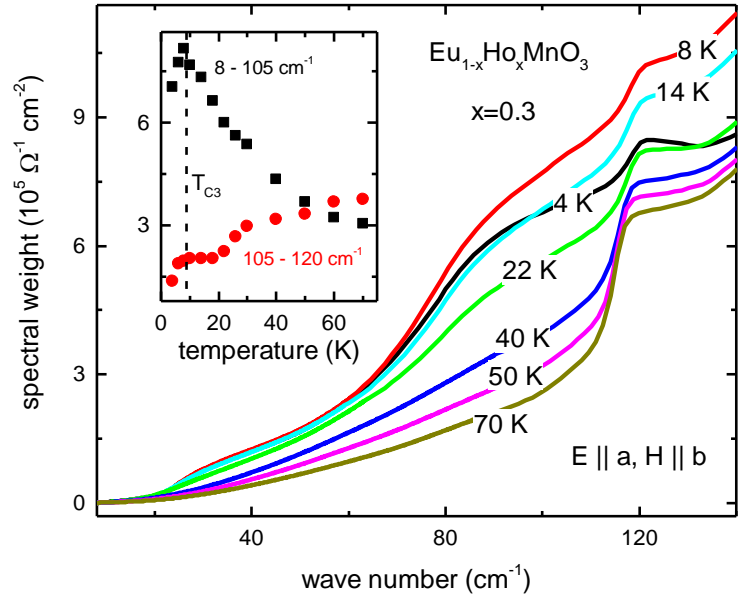

Fig. 9: (Color online) The spectral weight as function of wave number up to 140 cm$^{-1}$ in Eu$_{0.7}$Ho$_{0.3}$MnO$_3$ for $\mathbf{E}^\omega||\mathbf{a}$ and $\mathbf{H}^\omega||\mathbf{b}$ for temperatures between 4 K and 70 K. Inset: The integrated spectral weight as function of temperature. The electromagnon response corresponds to an integral from 8 to 105 cm$^{-1}$ (black), while the phonon mode is found in the integral from 105 to 120 cm$^{-1}$ (red). The phase transition temperature $T_{C3}$ is indicated by a dashed line.

**IV. Concluding remarks**

In this work we present a detailed THz and FIR study of Eu$_{1-x}$Ho$_x$MnO$_3$ for holmium concentrations x = 0.1 and 0.3. While similar systematic optical experiments on mixed multiferroic perovskites have been performed in the past[46,47] our study reveals new und rather unexpected results:

i) Well defined AFM resonances exist in the A-type AFM phase at 20.5 cm$^{-1}$. They can be excited with ac magnetic fields $\mathbf{H}^\omega||\mathbf{a}$ and $\mathbf{H}^\omega||\mathbf{c}$. These resonances soften on approaching the phase transition to the sinusoidally modulated collinear IC phase and exist only as weak and heavily damped modes in the IC AFM phase. Their eigenfrequencies seem to approach zero upon entering the paramagnetic phase. This behavior is expected as the eigenfrequencies should follow the temperature dependence of the sublattice magnetization.

ii) AFM resonances also exist in the helicoidal phase (Fig. 5), but their intensity approximately is by a factor of 5 lower compared to the resonance modes in the A-type phase. Nevertheless, AFM resonance modes in the helicoidal phases are still better defined than in the IC AFM phase. These modes significantly change their character when the spiral plane is switched with temperature: For $H^\omega||a$, a weak but narrow line is observed for the **bc** spiral at low temperatures which significantly broadens on entering the helical phase with the **ab** plane as the spiral plane.

iii) Crystal field excitations of the *4f* electrons of Ho were detected close to 55 cm$^{-1}$ and probably also close to 80 cm$^{-1}$. An excitation observed at 100 cm$^{-1}$ for x = 0.3 probably is a defect mode of the mixed crystal. Its temperature dependence of the intensity points rather towards a bosonic mode.

iv) Clear electromagnon response with the typical twin-peak structure of the perovskite manganites was observed for x = 0.3 in the helicoidal phases with **P**||**a** and **c**. The LF-EM is by a factor of 2 – 3 broader than the AFM resonance at a similar eigenfrequency. The width of the HF-EM is much broader and typically enhanced by a factor of 10 compared to classical antiferromagnetic resonances.

v) The electromagnon intensity is significantly reduced when the polarization switches from the **a** into the **c** direction. Interestingly, the optical weight of the electromagnons seems to be transferred completely from the lowest phonon mode for the multiferroic phase with **P**||**c**, but extra optical weight must be transferred from higher frequency phonon modes for the phase with **P**||**a**.

vi) Weak dynamic response close to 40 cm$^{-1}$ has been identified in the IC collinear phases for holmium concentrations x = 0.1 for $E^\omega||a$ and $H^\omega||b$ and **c** (Figs. 3 and 5). If this can be identified as electromagnon response is unclear. Its intensity is at least by a factor of 10 lower than the typical electromagnon response with $E^\omega||a$ in the helicoidal phases; its eigenfrequency is two times higher.

vi) Finally, these excitations which only exist in the IC phase are on top of an excitation continuum which also exists in the IC phase only and extends from the lowest frequencies at least up to 60 cm$^{-1}$. It could be related to two-magnon scattering processes which only exist in the collinear spin state, with the spins pointing along the **b** direction being sinusoidally modulated.

**Acknowledgements:**

This work has been supported by the Deutsche Forschungsgemeinschaft (DFG) via the Transregional Collaboration Center TRR 80: From Electronic Correlations to Magnetism (Augsburg, Munich, Stuttgart).